\documentclass[12pt]{iopart}

\usepackage{graphicx}
\usepackage{siunitx}
\usepackage{hyperref}

\newcommand{\epmsq}{~$e/\si{\micro\meter^2}$}

\begin{document}

\title[]{Probing surface charge densities on optical fibers with a trapped ion}

\author{
    Florian R. Ong,$^1$
    Klemens Sch\"uppert,$^1$
    Pierre Jobez,$^1$
    Markus Teller,$^1$
    Ben Ames,$^1$
    Dario A. Fioretto,$^1$
    Konstantin Friebe,$^1$
    Moonjoo Lee,$^2$
    Yves Colombe,$^1$
    Rainer Blatt,$^{1,3}$ 
    and Tracy E. Northup$^1$}

\address{$^1$Institut f\"ur Experimentalphysik,
    Universit\"at Innsbruck,
    Technikerstra\ss e 25,
    6020 Innsbruck,
    Austria}
\address{$^2$Department of Electrical Engineering,
    Pohang University of Science and Technology (POSTECH),
    Pohang, 37673,
    Korea}
\address{$^3$Institut f\"ur Quantenoptik und Quanteninformation,
    \"Osterreichische Akademie der Wissenschaften,
    Technikerstra\ss e 21a,
    6020 Innsbruck,
    Austria}
\ead{tracy.northup@uibk.ac.at}

\vspace{10pt}

\begin{indented}
\item[] \today
\end{indented}

\begin{abstract}

We describe a novel method
    to measure the surface charge densities on optical fibers
    placed in the vicinity of a trapped ion,
    where the ion itself acts as the probe.
Surface charges distort the trapping potential,
and when the fibers are displaced,
the ion's equilibrium position and secular motional frequencies are altered.
We measure the latter quantities for different positions of the fibers 
and compare these measurements to simulations 
in which unknown charge densities on the fibers are adjustable parameters.
Values ranging from $-10$ to $+50$\epmsq{ }were determined.
Our results will benefit the design and simulation of miniaturized experimental systems
    combining ion traps and integrated optics,
    for example,
    in the fields of quantum computation, communication and metrology.
Furthermore,
    our method can be applied to any setup
    in which a dielectric element can be displaced relative to a trapped charge-sensitive particle.

\end{abstract}

\section{Introduction}

Trapped-ion platforms have been used to demonstrate small prototypes 
for quantum computation \cite{Haeffner2008, Nielsen2010},
quantum simulation \cite{Blatt2012} and 
quantum communication \cite{Siverns2017}.
Micro-fabricated traps offer a promising route 
to scale up these prototypes into quantum devices 
capable of addressing classically intractable problems
\cite{Kielpinski2002, Hughes2011}.
However, along with the benefits of miniaturization come the challenges of preserving
the ion from perturbations from nearby surfaces,
as the ion acts as a sensitive probe of electric fields \cite{Maiwald2009, Huber2010, Narayanan2011}.
While stray DC-electric fields displace the ion and cause micromotion \cite{Berkeland1998},
the ion's motional coherence is easily disturbed by AC-electric field noise from trap surfaces
\cite{Brownnutt2015, Brama2012, Talukdar2016},
as quantified through heating rate measurements \cite{Brownnutt2015}. 
For metal surfaces,
these properties have been studied extensively \cite{Turchette2000, Daniilidis2011},
and methods have been developed to minimize disturbances \cite{Labaziewicz2008, Hite2012}.

In contrast, 
only a few studies have been undertaken on dielectrics close to trapped ions \cite{Harlander2010}.
However, the question of dielectric surface interactions with ions is highly relevant,
as the integration of optics into microtraps is expected 
to enable the scalable and efficient 
initialization, manipulation, and measurement of ion-based quantum bits
\cite{Monroe2013},
as demonstrated in recent proof-of-principle experiments,
e.g., with Fresnel lenses, waveguides, and high-finesse optical cavities
\cite{Jechow2011, Mehta2016, Cetina2013}.
Materials like indium tin oxide (ITO)
open up the possibility 
to produce transparent electrodes \cite{Eltony2013},
but they also introduce optical losses
precluding their use in high-finesse cavities.
Micro-fabricated cavities are especially interesting,
as these may be used both for classical optics applications and for cavity quantum electrodynamics (cavity QED) experiments,
which enable quantum interfaces between photons and ions
\cite{Steiner2013, Brandstaetter2013, Takahashi2017, VanDevender2010, VanRynbach2016}.
Research into surface charges on dielectrics dates back many decades \cite{Gross1949}
and has been both extensive and diverse
\cite{Imburgia2016, Othman2016}.
Unfortunately,
standard methods from this field
are not applicable to ion traps,
where in-situ measurements under ultra-high vacuum are required 
in order to probe the conditions experienced by trapped ions.

Here, we present a method
to estimate the surface charge densities
on exposed dielectric surfaces close to a trapped ion.
Our approach is similar to that
used for electric field measurements of metallic surfaces \cite{Narayanan2011}
but allows us to determine not only the corresponding electric fields
but also the surface charge densities.
Our system is a fiber-based cavity-QED setup, 
and we determine the surface charge densities on the facets 
and on the sides of the two optical fibers 
forming the cavity.
The principle of the presented method is the following: 
A charged object, here a fiber, is displaced in the vicinity of a trapped ion.
The potential experienced by the ion
is altered by the corresponding change of the static electric field.
We first measure both the ion's position and its secular frequency 
for different configurations of the charged optical fibers.
Then we simulate these quantities
using the unknown surface charge densities as adjustable parameters in the simulations
until an agreement with the data is found.
For our trap,
we need to include an additional so-called patch potential in the simulations 
which takes into account fixed stray charges on metallic trap surfaces \cite{Narayanan2011, Brownnutt2015, Eble2010, Xie2017}.

The reported work provides us with insights into our experimental system
    and has implications for the further development of quantum devices
    combining traps with integrated optics.
First,
    our method can be directly implemented in other setups
    where a dielectric element can be displaced in vacuum relative to an ion or,
    in principle,
    another trapped charge-sensitive particle.
Second,
    the values for the charge densities measured on the surfaces of our fibers,
    ranging from $-10$ to $+50$\epmsq,
    can serve as ballpark figures for researchers simulating quantum devices
    that have been designed to mitigate charging effects
    due to integrated optics.

In \sref{sec_setup},
we introduce our experimental setup.
\Sref{sec-meas} explains how the position and secular frequencies of a trapped ion are measured.
The simulation of these quantities is then described in \sref{sec-simul}.
In \sref{sec_patch}, we discuss how to measure patch potentials and how to account for their effects in simulations.
Our main results,
on characterizing a dielectric fiber near an ion,
are presented in \sref{sec_determining_charge}.
Here, we investigate two cases:
a positively charged fiber and an almost neutral fiber.
Finally,
in \sref{sec-discussion} we discuss the limits of our method
and consider
how it could be applied to other trapped-ion experiments.

\section{Experimental apparatus}
\label{sec_setup}

\begin{figure}
  \includegraphics[width=\textwidth]{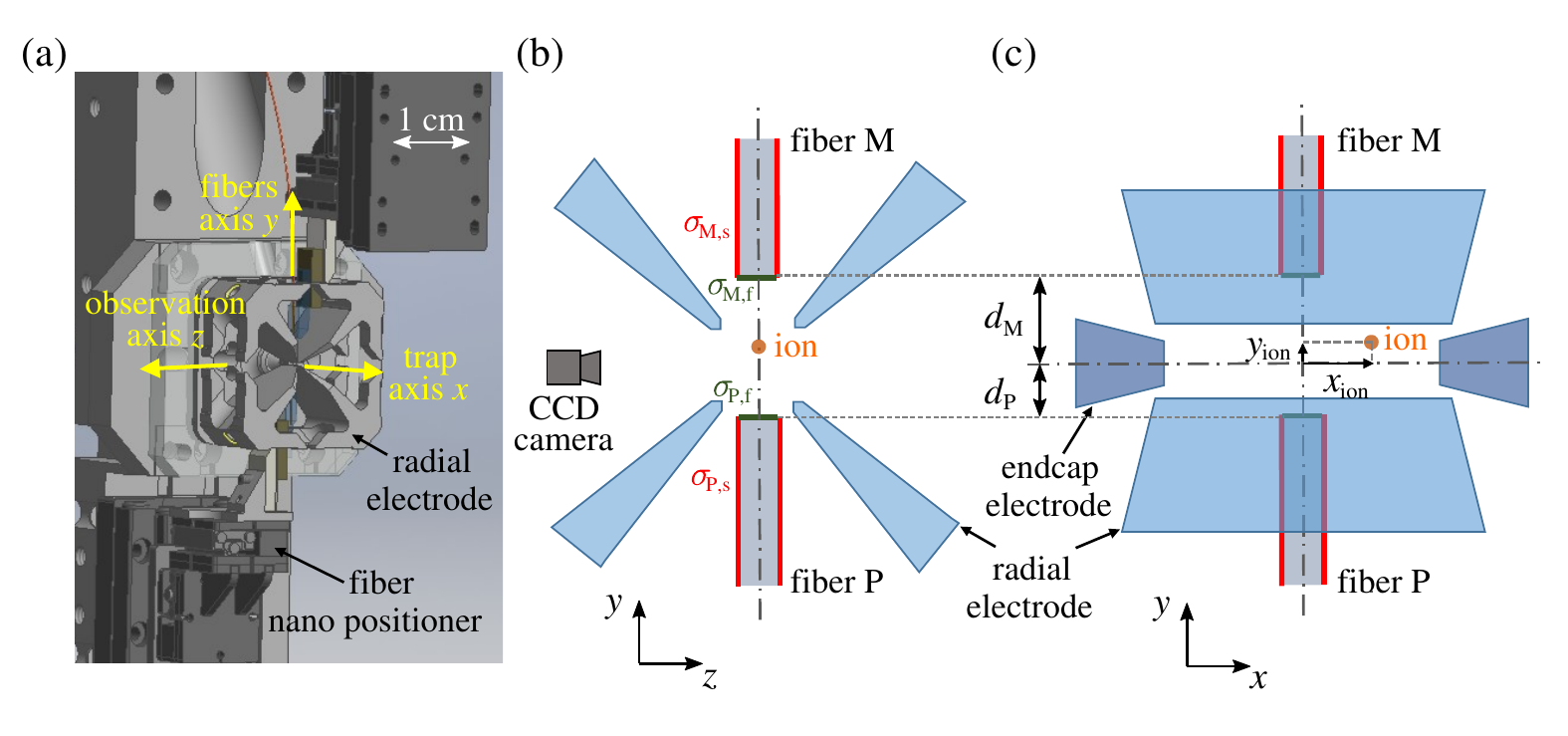}
  \caption{
      Experimental setup.
      (a) Cutaway drawing of the ion trap assembly.
      The linear Paul trap (center) is mounted on a sapphire base.
      The rest of the sapphire holder and one endcap electrode are not shown.
      Nanopositioner assemblies for the fibers are visible at the upper right and lower left.
      The upper fiber is shown,
      while the lower fiber has been omitted for simplicity.
      (b) Sketch, not to scale,
      of a view along the trap axis,
      indicating fibers M and P,
      surface charge densities $\sigma_{\rm M,f}$, $\sigma_{\rm M,s}$, $\sigma_{\rm P,f}$ and $\sigma_{\rm P,s}$, and the CCD camera position.
      (c) Sketch, not to scale,
      of the view from the camera,
      indicating the ion coordinates $x_\text{ion}$ and $y_\text{ion}$ and the ion--fiber distances $d_\mathrm{M}$ and $d_\mathrm{P}$.
  }
  \label{fig1}
\end{figure}

Our setup consists of a linear Paul trap \cite{Ghosh1996,Haeffner2008}
    combined with a fiber-based Fabry-P\'erot cavity \cite{Hunger2010}.
Two optical fibers 
    forming the cavity
    are inserted perpendicular to the trap axis
    according to the geometry described in \cite{Brandstaetter2013}.
The trap design,
    shown in \fref{fig1},
    is similar to the one reported in \cite{Guggemos2015}
    but with an ion-to-electrode distance of $\SI{270}{\um}$
    and a separation of \SI{5.1}{\milli\meter} between the endcap electrodes.
Compensation electrodes for minimizing ion micromotion are grounded for all measurements presented here.
The trap electrodes are made of gold-plated titanium 
and are mounted on a sapphire holder.
We apply a $\SI{30}{\mega\hertz}$ radio frequency (rf) signal on two opposing radial electrodes 
while grounding the remaining two radial electrodes.
Single $^{40}$Ca$^+$ ions are trapped 
at radial and axial secular frequencies of $\SI{1.6}{\mega\hertz}$ and $\SI{200}{\kilo\hertz}$, respectively.
Both optical fibers integrated with the ion trap 
can be independently translated relative to the ion-trap center 
along three axes
using in-vacuum nanopositioners\footnote{
    assemblies of SmarAct SL-0610-UHVT stages for the $x$- and $z$-axes and a SLC-1720-UHVT stage for the $y$-axis}.
\Fref{fig1}(a) shows a cutaway diagram of the ion trap 
in which most of the sapphire holder, one endcap, and one fiber have been removed. 

We define the origin of the coordinate system to lie at the trap center.  
The trap axis is along the $x$-axis,
and lies in the horizontal plane.
The fibers are mounted vertically, along the $y$-axis.
Although the fibers can be moved over a few \si{\milli\meter} along the $x$- and $z$-axes,
in the following measurements they are kept at fixed positions 
close to the origin with $x,z \in [-10,10]~\si{\micro\meter}$.
We use two different types of fibers:
fiber M is a multimode fiber\footnote{
    IVG Fiber Cu200/220}
with a cladding diameter of \SI{220}{\micro\meter}, 
and fiber P is a photonic crystal fiber\footnote{
    NKT Photonics LMA-20}
with a cladding diameter of \SI{230}{\micro\meter}. 
Both fibers' facets were machined into concave profiles with a CO$_2$ laser 
and were subsequently coated with a dielectric mirror stack of Ta$_2$O$_5$/SiO$_2$ bilayers,
with the outermost layer being SiO$_2$~\cite{Ott2016a}.
The photonic crystal fiber provides efficient incoupling to TEM$_{00}$ cavity modes,
whereas the multimode fiber is used for efficient outcoupling of light from the cavity~\cite{Hunger2010,Ott2016a}.
The protective copper and polyimide layers of fibers M and P, respectively,
were removed over more than \SI{10}{\milli\meter} behind the tips,
so that the side surfaces in the vicinity of the ion trap consist of the dielectric cladding.
The facets and the sides of the fibers thus constitute 
dielectric surfaces prone to trapping charges,
which may be positive or negative.
In this work,
we will assume that these trapped charges can be described 
by homogeneous surface charge densities,
called $\sigma_{\rm M,f}$, $\sigma_{\rm M,s}$, $\sigma_{\rm P,f}$ and $\sigma_{\rm P,s}$.
The labels M and P identify the fiber, and f and s stand for ``facet" and ``side".

\Fref{fig1}(b) presents a sketch (not to scale) of a view along the trap axis 
depicting the fibers' positions relative to the ion trap.
The facets of fibers M and P are held 
at distances $d_\mathrm{M}$ and $d_\mathrm{P}$, respectively, from the trap center.
In the absence of the fibers,
the ion is trapped at the trap center.
A charged fiber produces a static electric field that,
when the fiber is brought near the ion,
attracts or repels the ion.
Using a CCD camera,
we measure the corresponding horizontal shift $x_{\rm ion}$
and vertical shift $y_{\rm ion}$ of the ion in the $xy$ plane;
see \fref{fig1}(c).
The static electric field 
produced by a charged fiber 
also modifies the secular frequencies of the ion's motion in the trap,
which we also measure,
as we now describe.

\section{Measurement of ion position and secular frequency}
\label{sec-meas}

\begin{figure}
  \begin{center}
  \includegraphics[width=\textwidth]{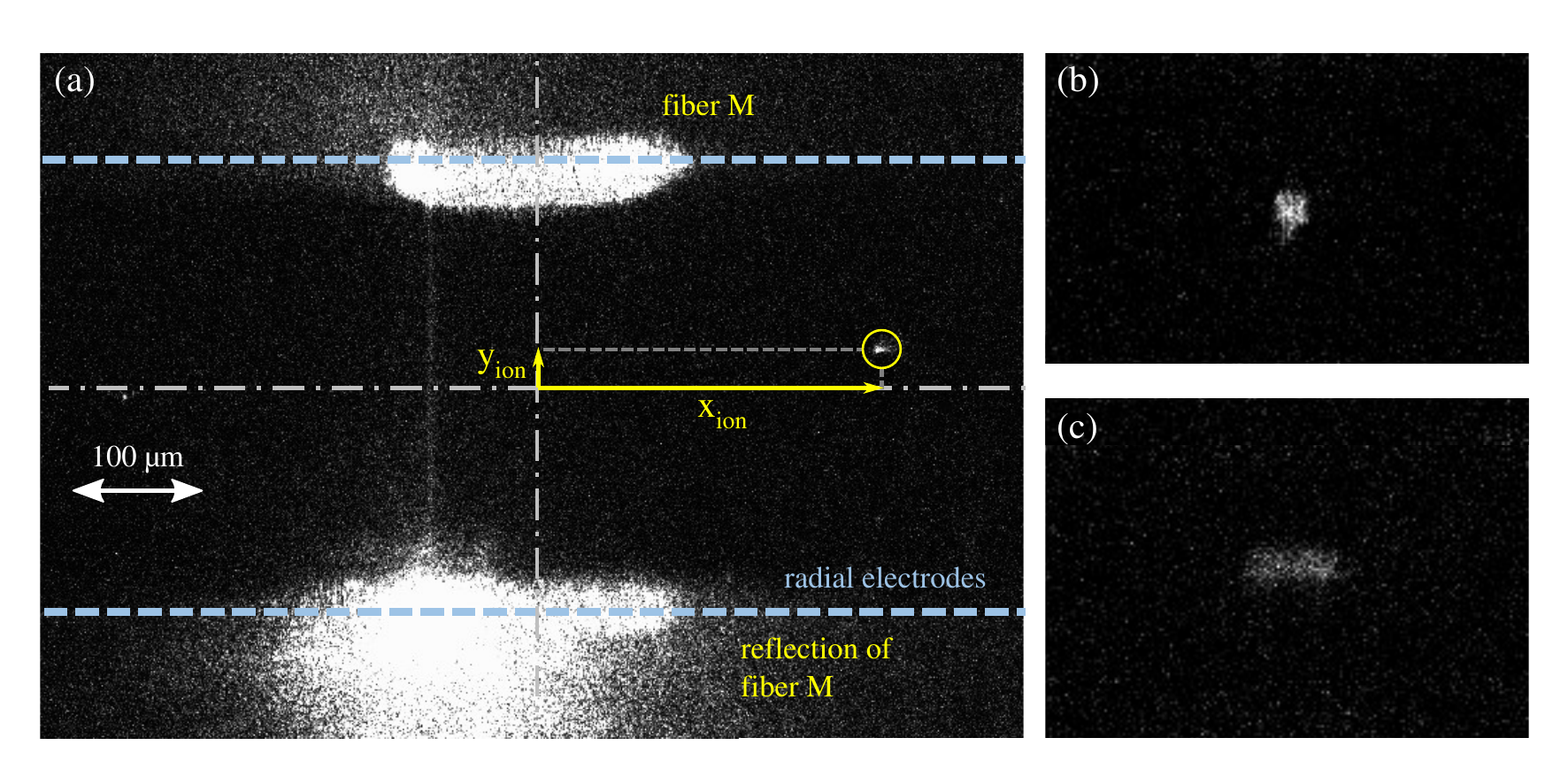}
  \caption{
  CCD camera images.
  (a) Full field of view, in which an ion (yellow circle) is displaced from the trap center in response to fiber M.
      The fiber tip is visible below the upper radial electrode.
      The lower electrode reflects scattered light from fiber M.
  (b) Close-up view of an unperturbed ion during a measurement of the axial secular trap frequency.
  (c) Close-up view of an ion that is resonantly driven at the axial secular trap frequency.
  }
  \label{fig2}
  \end{center}
\end{figure}

\Fref{fig2}(a) shows a CCD image of an ion displaced from the trap center in response to the presence of fiber M,
which in this example is positively charged.
The ion continuously fluoresces under illumination by a \SI{397}{\nano\meter} Doppler-cooling laser 
and an \SI{866}{\nano\meter} repumping laser \cite{Haeffner2008}.
Both fluorescence from the ion and scattered light
from the trap electrodes and inserted fibers 
pass through an objective and band-pass interferometric filters,
after which they are detected by the camera.

To infer the ion position $x_{\rm ion}$ along the trap axis,
we determine the center of the ion image from the CCD camera with a precision of $\pm1$ pixel.
The pixel to metric-distance conversion is calibrated using two independently known quantities,
the fibers' diameters and the separation of the radial electrodes,
yielding a conversion factor of  \SI{0.83(4)}{\micro\meter/px}.
The distances $d_\mathrm{M}$ and $d_\mathrm{P}$ between the fiber facets and the trap center 
are not directly measurable in our experiments.
Indeed,
for $d_\mathrm{M}, d_\mathrm{P}$ $> \SI{180}{\micro\meter}$,
which is the case for all data
presented in \sref{sec_determining_charge},
the fibers are shielded by the radial electrodes 
and thus not visible to the CCD camera.
In our setup,
the nanopositioners do not have closed-loop feedback capabilities,
and thus we do not have absolute position information about the fibers.
Instead, we rely on the reproducibility of the step-by-step nanopositioner displacements,
which we calibrate by displacing the fibers in the field of view of the CCD camera.
Single steps increasing or decreasing $d_\mathrm{M}$ 
correspond to displacements of 
\SI{1.7(1)}{\micro\meter} and \SI{2.2(1)}{\micro\meter},
respectively,
while single steps increasing or decreasing $d_\mathrm{P}$
correspond to respective displacements of
\SI{2.0(1)}{\micro\meter} and \SI{1.8(1)}{\micro\meter}
due to different loads of the nanopositioners.
We then assume that the distance travelled per step remains constant
when the fibers are retracted out of the field of view.
To test this assumption,
we start with a fiber at a reference position within the field of view.
We retract it by a given number of steps,
then reverse the direction by the number of steps 
that according to our calibration should cover the equivalent distance.
For fiber M, we recover the reference position within \SI{\pm 25}{\micro\meter} for any displacement up to \SI{2.0}{\milli\meter};
we thus estimate \SI{25}{\micro\meter} to correspond to a $2\sigma$ deviation 
and use a value of \SI{13}{\micro\meter} for the uncertainty in $d_{\rm M}$.
Fiber P has a reproducible behavior with the same uncertainty up to a retraction of \SI{0.75}{\milli\meter} from the trap center.
For larger displacements of fiber P up to \SI{1.6}{\milli\meter},
the load on the piezo nanopositioner is modified,
most likely due to contact between the fiber and the vacuum chamber,
resulting in a less reliable calibration.
In this regime,
we estimate the $2\sigma$ deviation of $d_{\rm P}$ to be \SI{60}{\micro\meter}
and the uncertainty to be \SI{30}{\micro\meter}.

We furthermore measure the axial secular trap frequency $\omega_{\rm ax}$.
We excite the axial mode of the ion's motion 
by applying a voltage of variable frequency $\omega_{\rm exc}$ to one of the endcap electrodes~\cite{Naegerl1998}.
Scanning $\omega_{\rm exc}$,
we are able to distinguish between an unperturbed ion, as in \fref{fig2}(b),
and an excited ion, as in \fref{fig2}(c), 
and to determine $\omega_{\rm ax}/2\pi$ within \SI{\pm2}{\kilo\hertz}.

We note that the same techniques could be applied 
to measure the ion position along the $y$-axis and the secular frequencies along both radial axes.
However, we do not present such measurements 
because we are not able to compare them with their simulated counterparts,
as will be explained in \sref{sec_patch}.

\section{Simulations of ion position and secular frequencies}
\label{sec-simul}

Simulations of the total potential energy landscape are performed
to estimate the ion's equilibrium position and secular frequencies 
for arbitrary trap biasing voltages
and for any charge state of the fibers according to the homogeneous surface-charge model described in \sref{sec_setup}.
The total potential is the sum 
of the rf pseudopotential \cite{Ghosh1996} and the static potentials 
generated by the endcap electrodes and by the charge densities on the fibers' surfaces.
We use finite element analysis software\footnote{
    COMSOL Multiphysics, AC/DC Module} 
to simulate the relevant electrostatic potentials. 
The simulations take into account the modification of the rf potential by the fibers
    by treating the fibers as dielectrics.
The simulated fibers are assigned a dielectric constant at rf equal to that of silica,
and with a length of \SI{6}{\milli\meter}.
The compensation electrodes are grounded in all cases.

We first simulate the static electric potential $U_{\rm rf}(\vec{r})$ generated 
by the voltage $V_0 = \SI{1}{\volt}$ applied to the rf pair of radial electrodes 
while all other electrodes are grounded and all surface charge densities on the fibers are set to zero.
The rf pseudopotential is then given by 
\[\Phi_{\rm rf}(\vec{r}) ~ = ~ \frac{Q^2}{4 M \Omega^2} \left( \frac{V_{\rm rf}}{V_0} \right)^2 \vec{\nabla}^2 U_{\rm rf}(\vec{r}),
\] 
where $M$ is the mass of a $^{40}$Ca$^+$ ion,
$Q$ is the elementary charge $e$,
and $\Omega$ and $V_{\rm rf}$ are the angular frequency and amplitude of the voltage applied to the rf electrodes. 
The potential energy due to the right or left endcap electrode $i \in\{R,L\} $,
when biased by a DC voltage $V_{\mathrm{ec},i}$, is given by 
\[
    \Phi_{\mathrm{ec},i}(\vec{r}) ~ = ~ Q  \left( \frac{V_{\mathrm{ec},i}}{V_0} \right) U_{\mathrm{ec},i}(\vec{r}),
\] 
where $U_{\mathrm{ec},i}(\vec{r})$ is the simulated electric potential generated 
when the voltage $V_0 = \SI{1}{\volt}$ is applied to the endcap $i$,
all other electrodes are grounded, and all surface charge densities are set to zero. 
Next, we calculate the potential energy $\Phi_{\sigma_{j,k}}$ 
due to a single fiber region with a surface charge density $\sigma_{j,k}$,
$j\in\{\mathrm{M,P}\}, k\in\{\mathrm{f,s}\}$.
This potential energy is given by
\[
\Phi_{\sigma_{j,k}}(\vec{r}) ~ = ~ Q  \left( \frac{\sigma_{j,k}}{\sigma_0} \right) U_{\sigma_{j,k}}(\vec{r}),
\]
where $U_{\sigma_{j,k}}(\vec{r})$ is the static electric potential simulated for  $\sigma_{j,k}=\sigma_0=1$\epmsq,
while all electrodes are grounded and all other surface charges densities are set to zero.
Finally, the total potential energy $\Phi_{\rm sim}(\vec{r})$ is given by 
\begin{equation}
    \Phi_{\rm sim}(\vec{r})=\Phi_{\rm rf}(\vec{r}) + \sum\limits_{i=\mathrm{R,L}} \Phi_{\mathrm{ec},i}(\vec{r}) + \sum\limits_{j=\mathrm{M,P} ; k=\mathrm{f,s} } \Phi_{\sigma_{j,k}}(\vec{r})\mathrm{.} \label{eq_total_potential}
\end{equation}

For each experimental configuration of the fiber distances $d_\mathrm{M}$ and $d_\mathrm{P}$ from the trap center,
we successively simulate $U_{\rm rf}(\vec{r})$, $U_{\mathrm{ec},i}(\vec{r})$, and $U_{\sigma_{j,k}}(\vec{r})$,
and we export a two-dimensional map of the corresponding electrical potentials in the $xy$ plane at $z=0$.
In separate post-processing software,
we build the total potential $\Phi_{\rm sim}(x,y)$ 
by summing the different contributions 
scaled by the voltages applied to the trap and for a given set of surface charge densities.

\begin{figure}
  \centering
  \includegraphics[width=\textwidth]{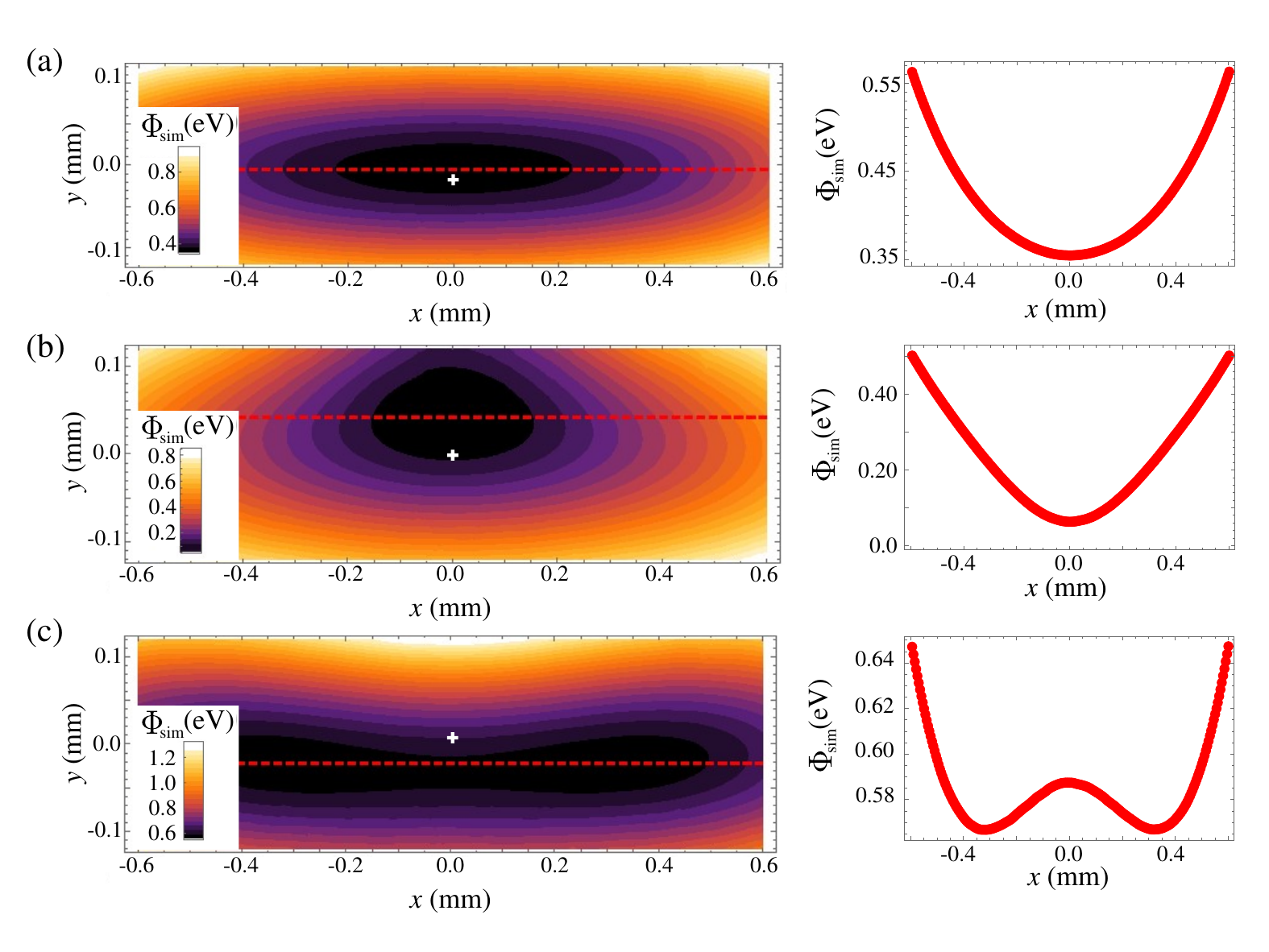}
  \caption{
      Comparison of simulated potential maps in the $xy$ plane 
      for different surface charges on fiber M while fiber P is held neutral.
      (a) $\sigma_{\rm M,f}=\sigma_{\rm M,s}= 0$,
      (b) $\sigma_{\rm M,f}=\sigma_{\rm M,s}= -4\sigma_0$,
      (c) $\sigma_{\rm M,f}=\sigma_{\rm M,s}= +4\sigma_0$, where $\sigma_0=1$\epmsq. 
      Line cuts are taken through the potential maps along dashed lines which intersect the potential minima and are plotted in the right column.
      Fiber M is located above the upper edge of these maps at a distance of $d_\mathrm{M} = \SI{0.5}{\milli\meter}$ from the trap center, while 
      fiber P is $d_\mathrm{P} = \SI{1.6}{\milli\meter}$ below the trap center
      depicted as a white cross.
  }
  \label{fig3}
\end{figure}

In \fref{fig3},
we show simulated total potentials for neutral, negative, and positive charge states of fiber M,
where $d_\mathrm{M} = \SI{0.5}{\milli\meter}$ and fiber P is held neutral.
The simulations are performed with parameters from our experimental setup:
    $V_{\rm rf} = \SI{111}{\volt}$,
    $\Omega = \SI{30.25}{\mega\hertz}$,
    $V_{\mathrm{ec},R} = V_{\mathrm{ec},L} = \SI{1.3}{\kilo\volt}$,
    and all other DC electrodes are grounded.
When fiber M is negative,
the ion is attracted by the fiber (\fref{fig3}(b)),
and its axial potential is stiffened,
that is, the curvature is steeper at the potential minimum.
On the other hand, when fiber M is positive, the ion is repelled.
In the example of \fref{fig3}(c), the repulsive force is so large 
that it splits the axial potential
generated by the endcaps
into a double-well,
as also observed in \cite{VanRynbach2016}.

For comparison with measured quantities,
we determine the axial equilibrium position and the axial secular frequency of the ion for each simulation.
As a starting point,
we take the mean position of all points for which the potential lies in the (lowest) first percentile.
For a circular area with a radius of \SI{50}{\micro\meter} around this minimum,
the potential is fitted with a two-dimensional parabolic curve
 so that the simulated potential around the ion's position can be approximated with a harmonic potential.
Translations along $x$ and $y$,
curvatures of the two semi-axes,
and a rotation of these semi-axes
are used as free parameters for this two-dimensional fit.
The ion position is given by the value for the translation along $x$,
and its axial secular frequency is extracted from the fitted curvature along the minor semi-axis.

\section{Trap imperfections and patch potential}
\label{sec_patch}

By simulating each experimental fiber configuration
using the methods of \sref{sec-simul},
one should be able to find a set of surface charge densities
$\{\sigma_{\rm P,f},\sigma_{\rm P,s},\sigma_{\rm M,f},\sigma_{\rm M,s} \}$ 
that reproduces the results of the measurements discussed in \sref{sec-meas}.

First, however,
the simulations should be validated independently of the fibers' influence.
To this end, 
the fibers were maximally retracted to minimize their influence on the ion.
With the left and right endcap voltages set to be equal, $V_{\rm ec,L} = V_{\rm ec,R}$,
the mean voltage $V_{\rm ec,M} = (V_{\rm ec,L} + V_{\rm ec,R})/2$ of the endcaps was tuned
to compare measurements of the axial secular trap frequency $\omega_{\rm ax}$ with the simulations.
The measured and the simulated data were fitted to the formula
    $\omega_{\rm ax} = \sqrt{\alpha V_{\rm ec,M} + \omega_0^2}$.
From this fit, 
we determined the factor $\alpha$,
which accounts for the geometry of the trap.
However, the factor determined from measurements,
$\alpha_{\rm exp} = \SI{2.3(1)}{\kilo\hertz^2/\milli\volt}$,
differed significantly from the factor determined from the simulation,
$\alpha_{\rm sim} = \SI{1.6(1)}{\kilo\hertz^2/\milli\volt}$.
This discrepancy can be explained by the margins of error in the trap fabrication.
In particular,
the geometric factor $\alpha$ is sensitive to the radial positions of the blades.
Adjusting the simulations 
so that the distances between the blades are larger by only about $7$\% reproduces measured data.
As we cannot determine the radial positions of the blades 
with uncertainties less than \SI{20}{\micro\meter},
we compensate for this discrepancy
by scaling the voltages
applied to the endcaps in our simulation correspondingly
to achieve the measured secular trap frequencies.

Furthermore,
the theory of our linear trap design does not include
an additional confinement $\omega_0$
in the expression above for $\omega_{\rm ax}$ \cite{Raizen1992}.
However,
for agreement with our measurements,
it was necessary
to include a positive confinement.
This confinement was found to be independent of the applied rf voltage in our experiment. 
Such a residual static potential is common in ion traps \cite{Brownnutt2015};
in our case, it is likely due to deposits of atomic calcium 
that have built up on radial electrodes after repeated loading of ions \cite{Eble2010, Narayanan2011, Xie2017}.
These patches deform the trapping potential and may additionally oxidize and subsequently trap charges,
thus contributing to the total electric potential.
In our case, the precise origins, locations, and shapes of these patches are unknown.
We heated the trap radial electrodes locally 
with strong laser light \cite{Gulde2003} with powers up to \SI{15}{\watt}.
After this treatment,
we observed a different confinement $\omega_0$.
We infer that we are able 
to manipulate these patches,
but this method does not allow us 
to control the patches precisely or to eliminate them.

These patches
significantly affect the equilibrium position of the ion and its secular frequency.
Since we aim 
to determine the charges on the fibers 
by precise measurement of these values, 
we need to understand quantitatively 
the role of this so-called patch potential.
 As the ion is displaced along the trap axis during our measurements,
we need to consider the perturbation along the trap axis
by estimating the patch potential $U_{\rm patch}(x)$.
We then add this estimation to the simulated potential,
such that the total potential including the patch effect along $x$ is given by $\Phi(x,y)=\Phi_{\rm sim}(x,y) + U_{\rm patch}(x)$. 

\begin{figure}
  \includegraphics[width=\textwidth]{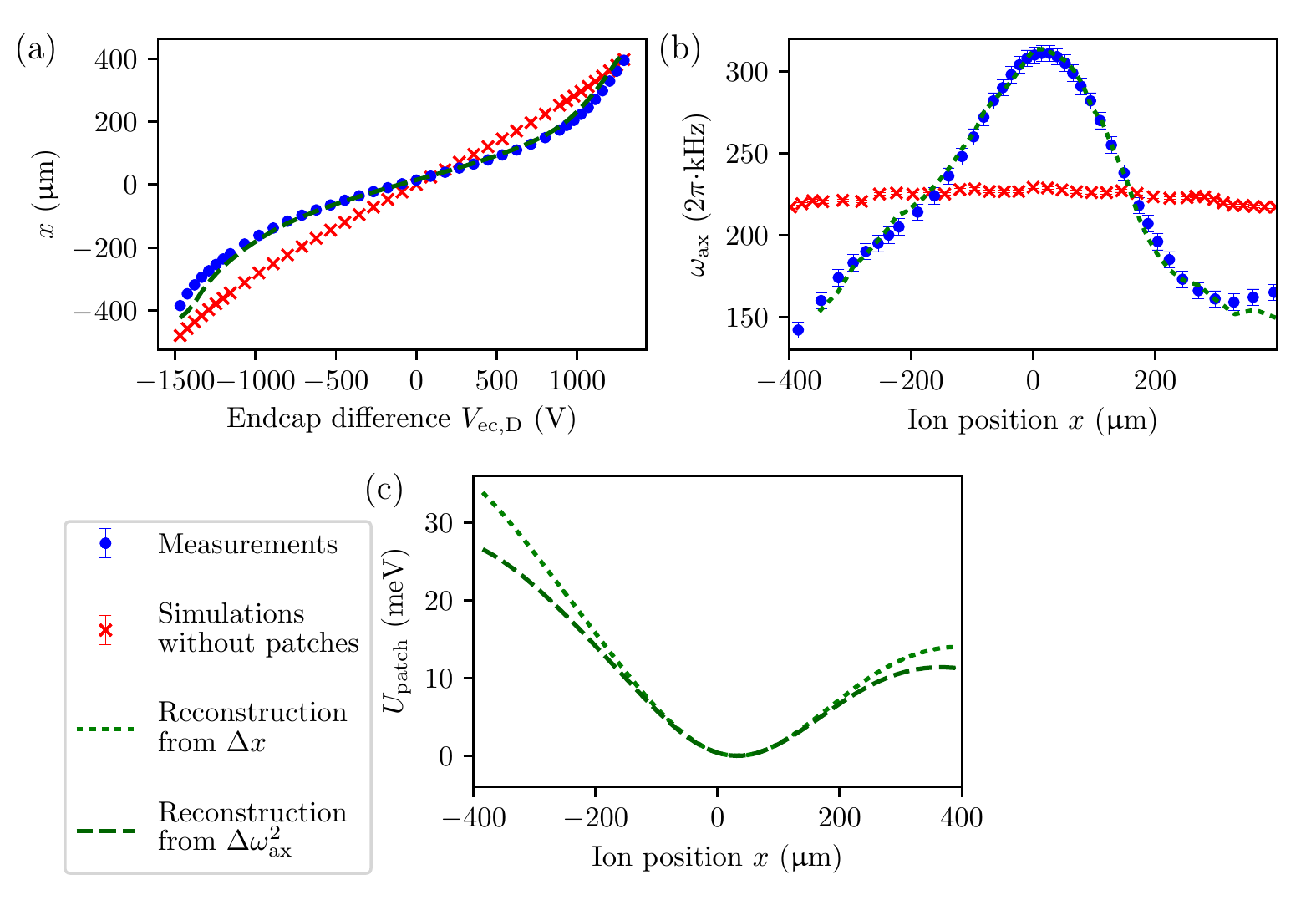}
  \caption{
    Reconstruction of the patch potential $U_{\rm patch}(x)$.
    (a) Measured and simulated ion positions $x$
      (blue points and red crosses, respectively)
      as a function of endcap voltage difference $V_\text{ec,D}$.
    (b) Measured and simulated axial frequencies
      (blue points and red crosses, respectively)
      as a function of ion position $x$.
    (c) Reconstructed patch potential from the discrepancies of the ion positions
      (green dotted line; see (a))
      and of the axial secular trap frequencies (green dashed line; see (b)).
    For the patch potential reconstruction based on trap frequencies,
    the corresponding ion positions have been simulated and plotted in (a) for comparison with the measured data,
    while for the patch potential reconstruction based on ion positions,
    the trap frequencies have been simulated and plotted in (b).
  }
  \label{fig4}
\end{figure}

To reconstruct $U_{\rm patch}(x)$,
we measure the ion position $x_{\rm exp}(V_{\rm ec,D})$ 
as a function of the endcap voltage difference $V_{\rm ec,D}=V_{\rm ec,L}-V_{\rm ec,R}$.
Here, we maintain a constant mean endcap voltage 
$V_{\rm ec,M}=\SI{1}{\kilo\volt}$.
The voltage difference $V_{\rm ec,D}$ is scanned
such that the ion moves over the full width of the CCD field of view (\SI{0.8}{\milli\meter}).
We then simulate the ion position $x_{\rm sim}$ in the absence of any patch potential.
For each voltage configuration $V_{\rm ec,D}$,
we assume that the measured total potential along the $x$-axis, $\Phi(x)$,
composed of the simulated potential $\Phi_{\rm sim}(x)$ and the patch potential,
can be approximated close to the ion by a harmonic potential:
\begin{equation}
    \Phi(x) = \Phi_{\rm sim}(x) + U_{\rm patch}(x) = \frac{1}{2} M \omega^2_{\rm exp} (x-x_{\rm exp})^2, 
    \label{eq_totxpotential}
\end{equation}
with the simulated harmonic potential around $x_{\rm sim}$
\begin{equation}
    \Phi_{\rm sim}(x) =  \frac{1}{2} M \omega^2_{\rm sim} (x-x_{\rm sim})^2.
\end{equation}
At the measured equilibrium position $x_{\rm exp}$,
the derivative of the total potential, $\Phi'(x_{\rm exp})$, vanishes:
\begin{equation}
    \Phi'(x_{\rm exp}) = \Phi'_{\rm sim}(x_{\rm exp}) + U'_{\rm patch}(x_{\rm exp}) = 0
\end{equation}
\begin{equation}
    U'_{\rm patch}(x_{\rm exp}) = - M \omega^2_{\rm sim} (x_{\rm exp}-x_{\rm sim})
\end{equation}
The position discrepancies between experiment and simulation for each measured voltage configuration $V_{\rm ec,D}$, $\Delta x(V_{\rm ec,D})=x_{\rm exp}(V_{\rm ec,D})-x_{\rm sim}(V_{\rm ec,D})$, are then used 
to recover $U_{\rm patch}(x)$ modulo a constant
by numerically integrating over the measured position $x_{\rm exp}(V_{\rm ec,D})$: 
\begin{equation}
    U_{\rm patch}(x) =  -M \omega_{\rm sim}^{2} \int_{0}^{x} \Delta x(V_{\rm ec,D}(x_{\rm exp})) dx_{\rm exp}.
\end{equation}

Alternatively, after differentiating equation~\eref{eq_totxpotential} twice,
\begin{equation}
    \Phi''(x) = \Phi''_{\rm sim}(x) + U''_{\rm patch}(x)
\end{equation}
\begin{equation}
    M \omega^2_{\rm exp} = M \omega^2_{\rm sim} + U''_{\rm patch}(x),
\end{equation}
we can use the discrepancy between the squared secular trap frequencies
\begin{equation}
    \Delta \omega^2(V_{\rm ec,D}) = \omega^2_{\rm exp}(V_{\rm ec,D}) - \omega^2_{\rm sim}(V_{\rm ec,D})
\end{equation}
to determine the curvature of the patch potential $U''_{\rm patch}$ at $x_{\rm exp}(V_{\rm ec,D})$.
After one integration over the position, 
the integration constant is chosen such
that the simulated equilibrium position without any endcap voltages ($V_{\rm ec,M}=0$) 
matches the one observed experimentally.
Following a second integration, we again end up with the reconstructed patch potential.

The two reconstruction methods produce patch potentials with nearly identical structure,
as shown in \fref{fig4}.
For both methods,
the starting point for integration is the point at 
which the simulated and measured ion positions are equal,
at $x=\SI{30}{\micro\meter}$.
Integration then proceeds outwards stepwise,
such that any errors will accumulate,
as can be seen in \fref{fig4}(c),
where the discrepancy between the two reconstruction methods is greatest at $x = \SI{\pm 400}{\micro\meter}$.
To understand whether these deviations are significant,
we extracted the axial positions and secular trap frequencies 
from the total simulated potentials $\Phi(x,y)=\Phi_{\rm sim}(x,y) + U_{\rm patch}(x)$
for both reconstructions.
For the reconstruction based on the secular trap frequency,
the corresponding axial position is plotted together with the measured and simulated ion position values in \fref{fig4}(a).
For the reconstruction based on ion position,
the corresponding secular trap frequency is plotted together with the measured and simulated values in \fref{fig4}(b).
Mean discrepancies of both positions and frequencies and of both reconstruction methods are on the order of 
measurement errors (\SI{0.8}{\um} and \SI{2}{\kilo\hertz}) 
and simulation errors (\SI{0.5}{\um} and \SI{2}{\kilo\hertz}),
indicating that variations in the reconstructed patch potentials seen in \fref{fig4}(c) are not significant.
In the following,
reconstructions are based on the mean of the two reconstructed patch potentials,
with which we observe a mean discrepancy of \SI{4.8}{\kilo\hertz}
compared to the measured secular trap frequencies with uncertainties of \SI{2}{\kilo\hertz}.
For the reconstructed ion position,
the error depends strongly on the actual position.
In the center the deviations are only about \SI{1}{\micro\meter}.
But for displacements larger than \SI{300}{\micro\meter} from the trap center,
the discrepancies between reconstruction and measurement are up to \SI{15}{\micro\meter}
due to errors accumulated by the integration for the reconstruction.
This characteristic is taken into account in further reconstructions 
by a position-dependent error.

We note that the reconstruction method described above depends on tuning the endcap voltages 
and is thus only valid along the trap axis.
We are unable to use the same approach to evaluate data on ion position and secular frequency measured along the radial axis $y$.
It is in principle possible to extend the reconstruction method to the radial direction,
since we can displace the ion along the $y$-axis using another set of static electrodes.
However, a systematic mapping of $U_{\rm patch}(x,y)$ in  the $xy$ plane is challenging 
because when the ion is displaced from the trap axis,
it is affected by micromotion and often fails to remain stable or trapped.
Therefore, we limit ourselves to a one-dimensional reconstruction of the patch potential along the trap axis.
Moreover, for our trap geometry,
the axial secular frequency is the lowest and thus the most sensitive to the fiber position,
independent of the sign of the fibers' surface charges.
As a result, information in the radial direction is not essential for estimating the charge state of the fibers,
although knowledge about $U_{\rm patch}(x,y)$ in two dimensions would help to refine this estimation.

\section{Determining the charge surface densities on the fibers}
\label{sec_determining_charge}

Finally, we turn to the determination of surface charge densities on the fibers.
We will present two cases,
both from our experimental setup,
that are particularly relevant for cavity-QED experiments 
in which an ion is trapped inside a fiber-based cavity.
First, we examine the situation of a fiber 
that was found to be positively charged 
after the system was first assembled,
evacuated, and used to trap ions.
Second, we discuss the case of an almost neutral fiber,
which offers favorable conditions for building short fiber cavities 
that result in minimal disturbance of the ion's position.
Both fibers are always present in the setup,
    but they are measured one after the other:
    while the fiber to be measured is displaced relative to the ion,
    the other fiber is maintained at a fixed distance.
The variations of position and secular frequency of the ion are thus 
    due to the moving fiber only.

\subsection{Positively charged fiber}
\label{subsec-posfiber} 

\begin{figure}
  \includegraphics[width=\textwidth]{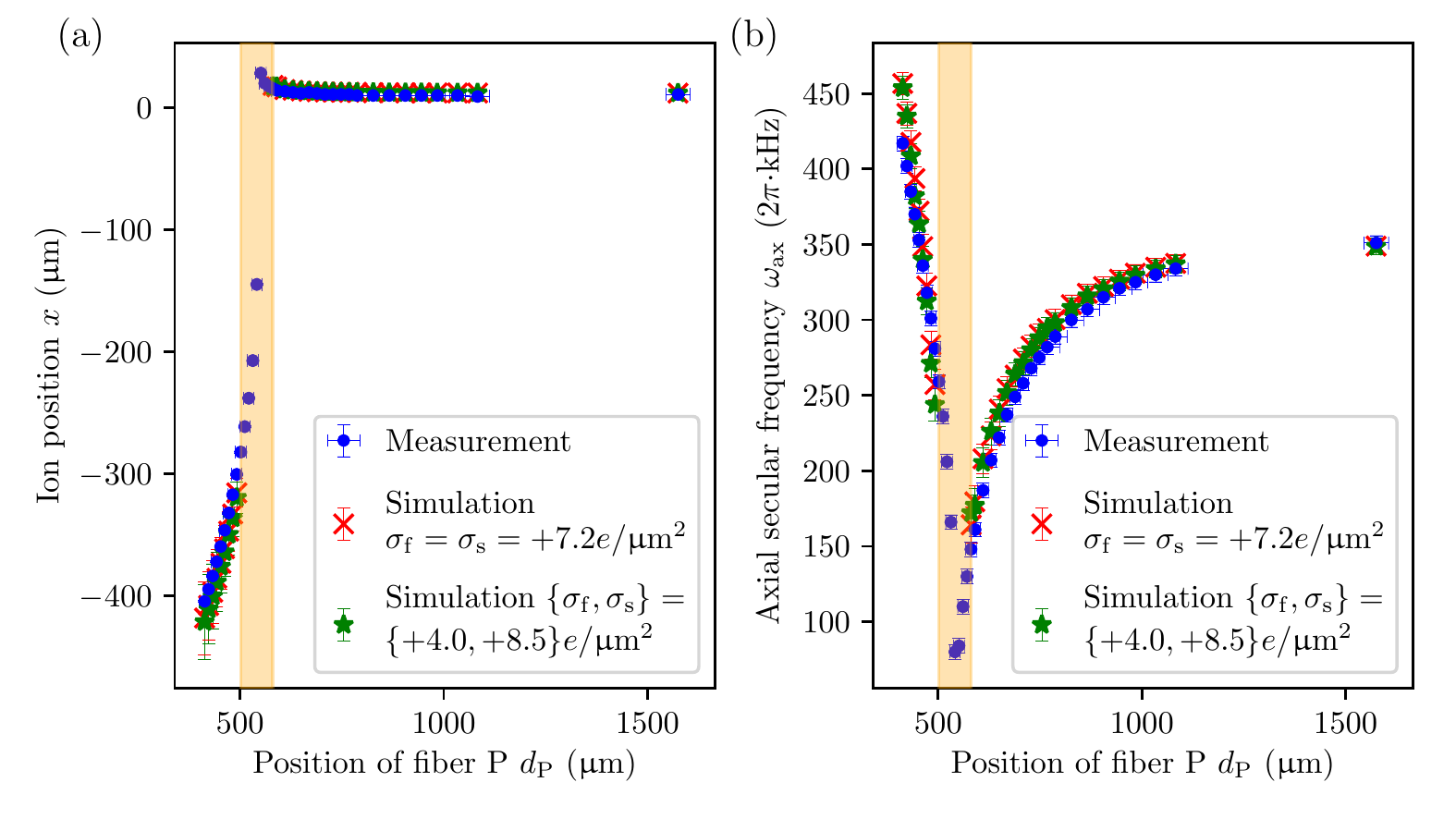}
  \caption{
      (a) Ion position $x_{\rm ion}$ and
      (b) axial frequency $\omega_{\rm ax}$
      as a function of $d_\mathrm{P}$, the position of fiber P,
      both measured (blue points)
      and simulated for 
      $\sigma_{\rm P,f}=\sigma_{\rm P,s}=+7.2$\epmsq{ }(red crosses)
      and for $\{\sigma_{\rm P,f}, \sigma_{\rm P,f}\}=\{+4.0, +8.5\}$\epmsq{ }(green stars).
      The orange shaded area indicates the transition 
      between a single-well and a double-well axial potential.
      The simulations were omitted in this region
      due to the anharmonic potential (see main text).
      Measurement error bars represent one standard deviation.
      Simulation error bars represent
      the uncertainties of the fit to the reconstructed potential 
      and the errors found for the reconstruction in \sref{sec_patch}.
  }
  \label{fig_measPos}
\end{figure}

\Fref{fig_measPos} shows 
the measured axial position and secular frequencies 
of the ion as a function of fiber P's position.
Here, fiber M remains retracted at $d_\mathrm{M}\approx$ \SI{2.0}{\milli\meter} 
and has no influence on the ion.
We verify this claim by translating fiber M to $d_\mathrm{M}\approx$ \SI{1.6}{\milli\meter}, 
which yields no measurable shifts in either the ion position or its axial frequency.
We conclude that fiber M's charges do not affect this measurement 
and set $\sigma_{\rm M,f}$ and $\sigma_{\rm M,s}$ to zero in the simulations.

As fiber P is brought towards the ion 
from a fully retracted position ($d_\mathrm{P}\approx \SI{1.6}{\milli\meter}$),
the axial frequency is first observed to decrease,
while the ion remains close to the trap center
but is slightly shifted away from the fiber.
At $d_\mathrm{P}\approx\SI{0.55}{\milli\meter}$,
the axial frequency reaches a minimum,
and as the fiber is moved closer to the trap center,
the axial frequency increases,
while the ion is now significantly displaced from the trap center.
This scenario corresponds to a positively charged fiber.
The static potential of the approaching positive fiber 
first relaxes the axial potential
that is provided by the endcaps,
and as the fiber is not exactly centered with respect to the ion, the fiber displaces the ion along the $x$-axis.
As the positively charged fiber is brought even closer to the trap center,
the repulsive force splits the axial potential into a double-well potential,
as illustrated in \fref{fig3}(c).
The transition between the single- and double-well cases 
happens at $d_\mathrm{P}\approx \SI{0.55}{\milli\meter}$.
For the double-well case,
either local minimum can be stably occupied in experiments.
In practice, the ion was reliably found on the side with $x<0$.

To determine the charge state 
$\{ \sigma_{\rm P,f}, \sigma_{\rm P,s} \}$
of fiber P,
the ion's positions and axial frequencies are simulated for the measured fiber's positions
and compared to the measurements.
By adjusting the charge distribution of the fiber used in the simulations,
the agreement between simulated and measured values
can be optimized,
that is,
the discrepancies between simulation and measurement for ion position and axial frequency
can be minimized.
For the purpose of this optimization,
the position and frequency discrepancies are normalized
so that a single value can be derived,
which is then minimized,
as we now describe.

We simulate the total potential 
$\Phi(x,y)=\Phi_{\rm sim}(x,y) + U_{\rm patch}(x) + \Phi_{\rm \sigma_{\rm P,s}}(x,y)+ \Phi_{\rm \sigma_{\rm P,f}}(x,y)$
for all measured positions of fiber P,
using the trapping voltages of the experiment.
In contrast to the applied voltages described in \sref{sec_patch},
here we raised the axial confinement
by applying a mean endcap voltage of $V_{\rm ec,M} = \SI{1.5}{\kilo\volt}$.
For the following simulations,
fiber P is positioned at $x=\SI{10}{\micro\meter}$,
which is within the uncertainty
with which we can position the fiber at the origin in our experimental setup
(cf. \sref{sec_setup}).
Simulations were carried out for fiber positions at $x=\{0, 5, 10, 15\}$~\si{\micro\meter} 
    and the best agreement between measurements and simulations was chosen.
From these reconstructed total potentials, 
we extract the simulated positions and axial frequencies.
We omit a reconstruction
around the transition region from a single to a double well
because a harmonic description is not valid in this regime;
see the orange shaded area in \fref{fig_measPos}.
For the double well,
we then selected the minimum in which the ion was observed.
For each simulated position and axial frequency,
we calculated the discrepancy to the measured value.
Each discrepancy was weighted
by the sum of the corresponding measurement, simulation, and reconstruction errors,
where all errors represent one standard deviation,
and where the error of the reconstructed position was considered to be position-dependent,
as described in \sref{sec_patch}.
This weighting serves to normalize the discrepancy,
so that a value of one means
that the simulation deviates by one standard deviation from the measurement.
With this normalization,
the squares of positions and axial frequencies at each fiber position
can be summed up to a single residual,
which is used 
to find the charge densities
$\{ \sigma_{\rm P,f}, \sigma_{\rm P,s} \}$
via a least-squares optimization code.

As the least-squares optimization code often converged to local minima
due to the noisy reconstruction,
residuals were calculated over a two-dimensional array of surface charge densities $\{ \sigma_{\rm P,f}, \sigma_{\rm P,s} \}$.
Here,
a correlation between the surface charge densities $\sigma_{\rm P,f}$ and $\sigma_{\rm P,s}$ was observed,
indicating that
removing charges on the fiber facet could be compensated for
by adding charges to the side,
or vice versa.
Nevertheless,
the global minimum was found over the 2D charge-density array. 
Using charge densities close to this global minimum as initial parameters,
the least-squares optimization code converged reproducibly to
$\sigma_{\rm P,f}=+4.0$\epmsq{ }and
$\sigma_{\rm P,s}=+8.5$\epmsq.
At this minimum,
a mean residual of $3.0$ for each measured point was calculated, 
meaning that the discrepancies between reconstruction and measurement 
were three times larger than the errors expected 
from the measurement, simulation and reconstruction.
Positions and frequencies corresponding to these optimized surface-charge values
are plotted in \fref{fig_measPos}.
This analysis was also carried out 
for a single charge density on the fiber's surface,
that is,
for the constraint $\sigma_{\rm f} = \sigma_{\rm s}$.
A value of $\sigma_{\rm f}=\sigma_{\rm s}=7.2$\epmsq{ }was found with a mean
residual of $3.1$ standard deviations.
This result underscores the range of possible surface charge densities
which are able to reproduce the measured data with similar agreement.

\subsection{Almost neutral fiber}
\label{subsec-neutral}

In coupling a trapped ion to an optical cavity, one has to take into account the electric field generated 
by any surface charges on the cavity mirrors, and it is generally desirable to minimize the effect of this field on the trapping potential.
In this regard, fiber-based Fabry-P\'erot cavities 
present a particular challenge
due to the small distances (typically a few hundred microns) between ion and dielectric mirror surfaces.
In our setup,
trapped ions were highly sensitive
to charges on the fibers
due to the low frequency of axial confinement (see \sref{sec_setup}).
Ideally, one would like to work without charges on the fibers,
but in practice this is challenging.
A viable alternative consists of obtaining distributions of positive and negative charges 
that almost cancel each other's effects at the location of the ion. 

\begin{figure}[ht]
  \includegraphics[width=\textwidth]{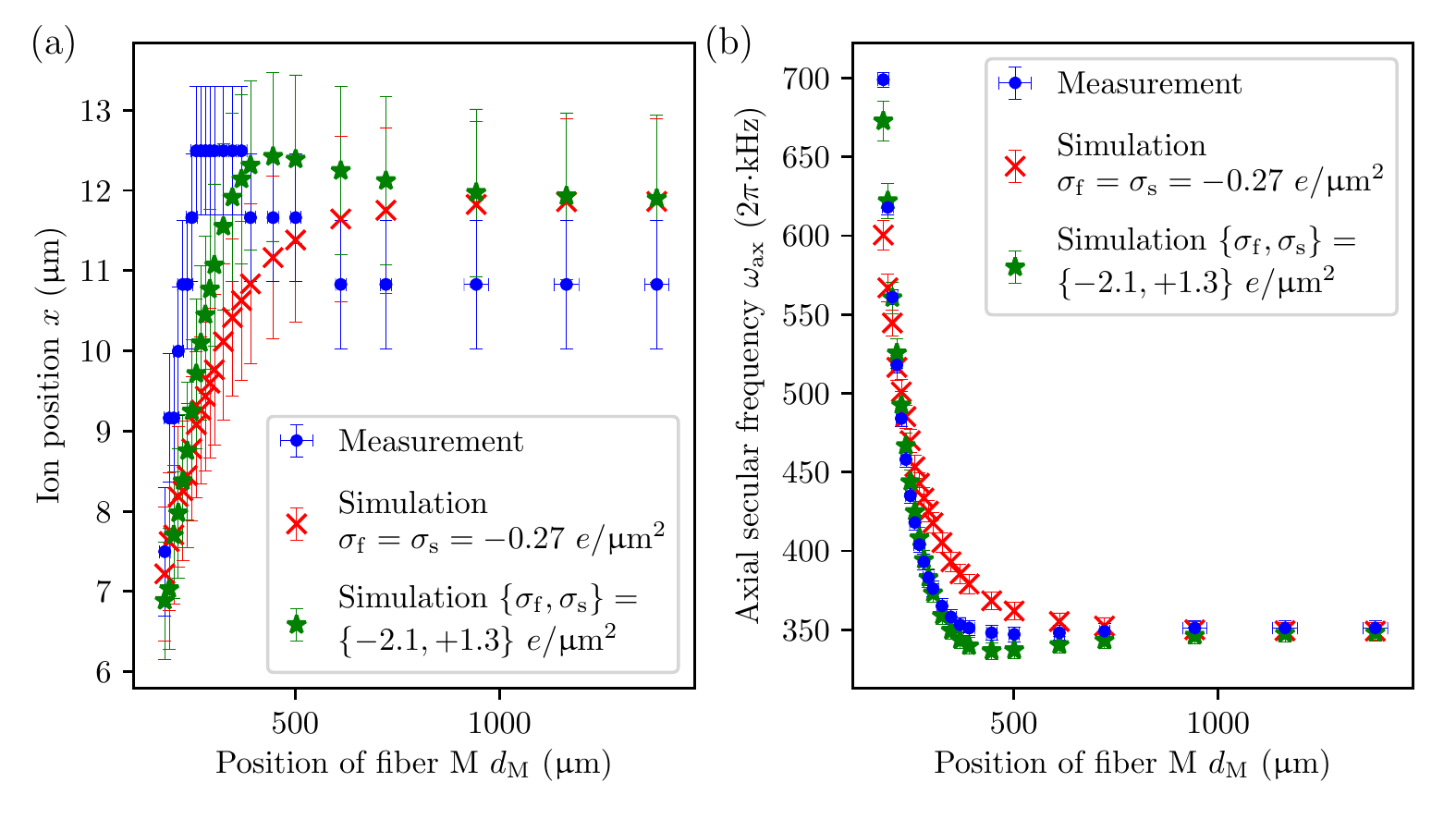}
  \caption{
      (a) Ion position $x_{\rm ion}$ 
      and (b) axial frequency $\omega_{\rm ax}$ 
      as a function of $d_\mathrm{M}$, the position of fiber M,
      both measured (blue points)
      and simulated for 
      $\sigma_{\rm P,f}=\sigma_{\rm P,s}=-0.27$\epmsq{ } (red crosses)
      and for
      $\{ \sigma_{\rm M,f}, \sigma_{\rm M,s}\}= \{ -2.1, +1.3\}$\epmsq (green stars).
      Measurement error bars represent one standard deviation.
      Simulation error bars represent the uncertainties of the fit to the reconstructed potential 
      and the errors found for the reconstruction in \sref{sec_patch}.
  }
  \label{fig_measNeg}
\end{figure}

We were able to modify the charge distribution of fiber M to achieve such a cancellation; the method will be discussed briefly in \sref{sec-discussion}.
In \fref{fig_measNeg},
the blue points show the measured axial positions and axial frequencies of the ion as a function of $d_\mathrm{M}$ 
while fiber P is kept retracted at $d_\mathrm{P} \approx \SI{1.6}{\milli\meter}$.
When fiber M is brought towards the trap center,
the ion remains close to the trap center but is first shifted slightly away from the fiber, then slightly towards it.
Its axial frequency increases, 
indicating that the ion sees a negative charge distribution overall.
Moreover, in contrast to the situation described in \sref{subsec-posfiber},
fiber M can be brought within \SI{200}{\micro\meter} of the ion
while the ion remains within \SI{4}{\micro\meter} of the trap center.

Following the approach described for the positively charged fiber,
we optimize the parameters $\{ \sigma_{\rm M,f}, \sigma_{\rm M,s} \}$ 
to obtain the best agreement between simulations and data.
Here,
fiber M's position was set to $x=\SI{5}{\micro\meter}$
following simulations over a range of positions within the experimental positioning uncertainty,
as explained in \sref{subsec-posfiber}.
In the simulations, for the retracted fiber P, we set
$\{ \sigma_{\rm P,f}, \sigma_{\rm P,s} \} = \{ +4.0,+8.5 \}$\epmsq, as determined previously.
The least-squares optimization code converges to 
surface charge densities of
$\{ \sigma_{\rm M,f}, \sigma_{\rm M,s} \} = \{ -2.1,+1.3\}$\epmsq 
{ }for fiber M
with a residual mean discrepancy of $1.5$.
Again, the reconstruction was also optimized for a single surface charge density.
Here, a minimum mean residual of $4.9$ standard deviations was found
for $\sigma_{\rm f}=\sigma_{\rm s}= -0.27$\epmsq.
The significantly higher discrepancy
between simulation and experiment in this case indicates
the need for different surface charge densities 
on the fiber's facet and side 
to account for the measured data properly.

When fiber M is close to the trap center,
the ion is mostly influenced by the negative charges on the fiber facet
and reacts accordingly.
But when fiber M is far away,
the positive and negative charges roughly compensate for one another,
perturbing the ion minimally:
Only when $d_\mathrm{M}$ is less than \SI{0.5}{\milli\meter} does the ion start 
to respond via its position and axial frequency to a displacement of the fiber.
In contrast, small displacements of fiber P can be detected via shifts of the ion's axial frequency 
even when the fiber is fully retracted ($d_\mathrm{P} \approx \SI{1.6}{\milli\meter}$).

\section{Discussion}
\label{sec-discussion}

We have described a method 
to model the surface charge densities on optical fibers next to a trapped ion in a way 
that reproduces experimental results.
In this work, we have shown a detailed analysis 
for one patch potential configuration in \sref{sec_patch}
and for two surface charge distributions of fibers in \sref{sec_determining_charge}.
Furthermore,
following these methods,
for several additional charge states of the fibers and patch potentials on the radial electrodes,
we were able to estimate a set of charge densities ranging from $-10$ to $+50$\epmsq.
This result could be of interest to designers of ion traps
    combining integrated optics and using similar materials.
Indeed,
    the quantity of charges trapped on dielectric surfaces is often unknown and uncontrollable.
Thus,
    the consideration of a realistic range could benefit the design and simulation 
    of mitigation strategies against the detrimental effects of these charges.

Comparing measurements of the ion's position and its axial secular frequency,
charge densities were determined via least-squares optimization,
but
the reconstructions for the two fibers had final mean discrepancies of
$1.5$ and three standard deviations
with respect to the measurements.
Two main limitations, described below,
prevent us from determining the charge densities with better 
agreement between reconstruction and measurement.

\subsection{Limits of the method}

First, the model we use to describe how the charges are distributed over the fibers does not account for complex structures: 
We assume that there is a homogeneous distribution of charges across the fiber facet and a second homogeneous distribution along the fiber side.
This choice is dictated by simplicity and symmetry in the absence of better knowledge of the actual charge distribution.
An even simpler choice would have been a single homogeneous distribution over the complete fiber surface.
This would have worked reasonably well for the positive fiber P,
as seen in \fref{fig_measPos},
but for fiber M, our two-parameter approach was essential to account for the data,
as seen in \fref{fig_measNeg}.
One could think of refining the model by dividing the exposed area into additional zones,
each carrying a different homogeneous density, or by parametrizing non-homogeneous distributions.
But without physical justification
to guide these choices,
one would end up with a wide parameter space to optimize,
which would be time- and resource-consuming.

Second, we suffer from imperfect knowledge of the patches described in \sref{sec_patch}.
Although we were able to gain information 
that allowed us to account for the effect of the patch along the $x$-axis,
this reconstruction is limited, 
particularly because we were not able to characterize the patch potential along the radial axis~$y$.
Not only could we have used measurements of the same quantities in the radial direction for a reconstruction,
but also we could have combined the analyses of measurements 
 and simulations along the two axes
to gain more precision of the determined surface charge densities.

\subsection{Temporal stability of charges}

The charge distributions in our setup shift over time,
affecting the timescales over which measurements are reproducible.
Here, we discuss those timescales and how they are accounted for in our measurements.

Recall that in \sref{subsec-neutral}, the charge distribution of fiber M had been modified to achieve a near-neutral state.  This modification was the result of a laser-based procedure that we have recently developed and investigated in which 
photoelectrons are created 
by the illumination of trap electrodes with around \SI{30}{\milli\watt} of \SI{393}{\nano\meter} laser light \cite{Ames2019}.
During this illumination, these electrons can be influenced
by applied voltages on particular electrodes,
after which the charge states on both the fibers and patches are observed to shift.
The patch potentials on the electrodes are found 
to evolve on a time scale of a few minutes, stabilizing after about one hour,
presumably due to charge rearrangement.
The charge state of the fibers
initially stabilizes within a few hours,
but continues to drift on another,
much slower time scale.

This procedure provides us with the flexibility 
to tune the charge state of the fibers 
to be either more negative or more positive.
However, one shortcoming is that the charges on the fibers continue to drift.
Fiber P reaches its final charge state within about one week,
whereas fiber M needs about one day.
These different time scales may be 
due to the coatings of the fibers,
which are more than \SI{10}{\milli\meter} behind the fiber facet
(\sref{sec_setup}):
Fiber P has an insulating coating, 
whereas fiber M has a metallic coating,
which could shorten the time scale of charge rearrangement.
Moreover,
if the fibers are translated via their nanopositioning stages 
before the final charge state has been reached,
the charge on the fibers shifts.  
Experimentally,
we address this 
by waiting for the charge state of both fibers to stabilize 
before carrying out measurements such as those 
presented in \fref{fig_measPos} and \fref{fig_measNeg}.
We have verified that after a sufficient waiting time,
translation of the fibers does not affect the measurement results within error bars.

\subsection{Possible applications}

In spite of these limitations,
our method provides a reconstruction of the patch potential along the ion-trap axis
and an estimation of the surface charge density of dielectrics.
This method can be replicated in any setup
    where a dielectric object can be displaced relative to a trapped ion or another charge-sensitive particle.
However,
    it cannot be applied to measure the charge density of a fixed object.
Only by displacing a charged dielectric object
    can the ion experience a change of electric field
    which could be attributed to the displaced charges.
Displacing an ion relative to a fixed dielectric object can only be used
    to measure the electric field generated by this object,
    as described in \cite{Narayanan2011}.

In cavity QED setups involving trapped ions and optical cavities based on dielectrics,
building short cavities to maximize the ion-cavity coupling rate is challenging 
because the mirrors tend to carry charges.
The ability to assess the charge densities on the fibers and their evolution 
will aid in the development of procedures to discharge the fibers.
Furthermore, knowledge of the typical charge densities 
that can be expected under a given set of conditions
will enable targeted approaches to improve experimental setups,
such as by incorporating elements to reduce the adverse effects of charges on trapped ions.
These considerations also apply to cavity QED setups using Rydberg atoms 
due to their large sensitivity to electric fields \cite{Guerlin2010, Wade2016}.

Our method is relevant for the field of trapped ions in general,
where systems designed for scalable quantum computing or quantum simulation 
can benefit from component miniaturization.
In a microfabricated trap,
ions can be stored a few tens of micrometers away from the trap surfaces,
which may generate spurious electric fields \cite{Hughes2011,Siverns2017}.
Such fields have been found to be correlated to ion heating \cite{Daniilidis2011}.
Efforts are also underway to integrate optical elements required 
to address ions with laser beams \cite{Mehta2016},
to collect ions' fluorescence \cite{Jechow2011, Eltony2013, VanDevender2010},
or to build fiber-based cavities \cite{Brandstaetter2013, Ragg2019}.
While these elements are fixed with respect to the ion trap, more complex future tasks, such as selective addressing or imaging,
may benefit from movable mirrors, lenses, fibers, or waveguides placed close to ions.
Understanding the charge state of these elements is desirable in order to minimize detrimental effects of the corresponding parasitic fields,
such as trapping instability, excess micromotion or anomalous heating, 
in view of optimizing the quantum operations for which a given device was designed.
More broadly, our method can be valuable for any experiment 
where charging effects or local electric fields on surfaces need 
to be understood or mitigated.

\section{Conclusion}

We have presented a method 
to determine the surface charge density on optical fibers 
placed in the vicinity of a trapped ion, assuming a simplified model of two homogeneous surface charge areas per fiber.
This method relies on 
measuring the ion's position and its secular frequency
for different positions of the fibers,
and on comparing these measurements to simulations 
with the unknown charges densities as adjustable parameters.
In our experimental setup,
the situation is complicated by the presence of additional confinement likely due to calcium depositions on the trap radial electrodes.
Following reconstruction of the corresponding patch potentials,
simulations were found to be in agreement with measurements,
allowing us to determine fiber surface-charge densities.

We expect our approach to allow for better characterization of
dielectric surfaces close to trapped ions.
Scalable quantum-computing devices making use 
of micro-fabricated traps and integrated optics 
may benefit from the capability
to assess the charge state of dielectric components
placed in the vicinity of ions.
In particular,
cavity-QED setups using fiber-based cavities
could be improved with the knowledge about the surface charge densities.
With the presented laser-based procedure
and the method to determine the surface charge densities,
we were able to reduce the charges on our fibers
so that we could trap an ion inside the fibers
aligned for cavity-QED experiments.

\section*{Acknowledgements}

We thank J. Ghetta and C. Roos for helpful discussions and insights.
This project has received funding 
from the European Union's Horizon 2020 research and innovation programme under the Marie Sk\l{}odowska-Curie Grant Agreement No. 656195;
from the Austrian Science Fund (FWF) through the SFB FOQUS, Project F 4019, and the Elise Richter Programme, Project V 252;
and from the Army Research Laboratory's Center for Distributed Quantum Information, Cooperative Agreement No. W911NF15-2-0060.

\section*{References}

\bibliographystyle{iopart-num}
\bibliography{charge_sensing}

\providecommand{\newblock}{}
\begin{thebibliography}{10}
\expandafter\ifx\csname url\endcsname\relax
  \def\url#1{{\tt #1}}\fi
\expandafter\ifx\csname urlprefix\endcsname\relax\def\urlprefix{URL }\fi
\providecommand{\eprint}[2][]{\url{#2}}

\bibitem{Haeffner2008}
H\"affner H, Roos C~F and Blatt R 2008 {\em Phys. Rep.\/} {\bf 469} 155--203
  ISSN 0370-1573 \urlprefix\url{https://doi.org/10.1016/j.physrep.2008.09.003}

\bibitem{Nielsen2010}
Nielsen M~A and Chuang I~L 2010 {\em Quantum Computation and Quantum
  Information\/} (Cambridge: Cambridge University Press) ISBN 9781107002173
  \urlprefix\url{https://isbnsearch.org/isbn/9781107002173}

\bibitem{Blatt2012}
Blatt R and Roos C~F 2012 {\em Nat. Phys.\/} {\bf 8} 277--284 ISSN 17452473
  \urlprefix\url{https://doi.org/10.1038/nphys2252}

\bibitem{Siverns2017}
Siverns J~D and Quraishi Q 2017 {\em Quantum Inf. Process.\/} {\bf 16} 314--356
  ISSN 1573-1332 \urlprefix\url{https://doi.org/10.1007/s11128-017-1760-2}

\bibitem{Kielpinski2002}
Kielpinski D, Monroe C and Wineland D~J 2002 {\em Nature\/} {\bf 417} 709--711
  ISSN 0028-0836 \urlprefix\url{https://doi.org/10.1038/nature00784}

\bibitem{Hughes2011}
Hughes M~D, Lekitsch B, Broersma J~A and Hensinger W~K 2011 {\em Contemp.
  Phys.\/} {\bf 52} 505--529
  \urlprefix\url{https://doi.org/10.1080/00107514.2011.601918}

\bibitem{Maiwald2009}
Maiwald R, Leibfried D, Britton J, Bergquist J~C, Leuchs G and Wineland D~J
  2009 {\em Nat. Phys.\/} {\bf 5} 551--554
  \urlprefix\url{https://doi.org/10.1038/nphys1311}

\bibitem{Huber2010}
Huber G, Ziesel F, Poschinger U, Singer K and Schmidt-Kaler F 2010 {\em Appl.
  Phys. B\/} {\bf 100} 725--730 ISSN 1432-0649
  \urlprefix\url{https://doi.org/10.1007/s00340-010-4148-x}

\bibitem{Narayanan2011}
Narayanan S, Daniilidis N, M\"oller S~A, Clark R, Ziesel F, Singer K,
  Schmidt-Kaler F and H\"affner H 2011 {\em J. Appl. Phys.\/} {\bf 110} 114909
  ISSN 00218979 \urlprefix\url{https://doi.org/10.1063/1.3665647}

\bibitem{Berkeland1998}
Berkeland D~J, Miller J~D, Bergquist J~C, Itano W~M and Wineland D~J 1998 {\em
  J. Appl. Phys.\/} {\bf 83} 5025--5033
  \urlprefix\url{https://doi.org/10.1063/1.367318}

\bibitem{Brownnutt2015}
Brownnutt M, Kumph M, Rabl P and Blatt R 2015 {\em Rev. Mod. Phys.\/} {\bf 87}
  1419--1482 \urlprefix\url{https://doi.org/10.1103/RevModPhys.87.1419}

\bibitem{Brama2012}
Brama E, Mortensen A, Keller M and Lange W 2012 {\em Appl. Phys. B\/} {\bf 107}
  945--954 ISSN 1432-0649
  \urlprefix\url{https://doi.org/10.1007/s00340-012-5091-9}

\bibitem{Talukdar2016}
Talukdar I, Gorman D~J, Daniilidis N, Schindler P, Ebadi S, Kaufmann H, Zhang T
  and H\"affner H 2016 {\em Phys. Rev. A\/} {\bf 93} 043415
  \urlprefix\url{https://doi.org/10.1103/PhysRevA.93.043415}

\bibitem{Turchette2000}
Turchette Q~A, Kielpinski, King B~E, Leibfried D, Meekhof D~M, Myatt C~J, Rowe
  M~A, Sackett C~A, Wood C~S, Itano W~M, Monroe C and Wineland D~J 2000 {\em
  Phys. Rev. A\/} {\bf 61} 063418
  \urlprefix\url{https://doi.org/10.1103/PhysRevA.61.063418}

\bibitem{Daniilidis2011}
Daniilidis N, Narayanan S, M\"oller S~A, Clark R, Lee T~E, Leek P~J, Wallraff
  A, Schulz S, Schmidt-Kaler F and H\"affner H 2011 {\em New J. Phys.\/} {\bf
  13} 013032 \urlprefix\url{https://doi.org/10.1088/1367-2630/13/1/013032}

\bibitem{Labaziewicz2008}
Labaziewicz J, Ge Y, Antohi P, Leibrandt D, Brown K~R and Chuang I~L 2008 {\em
  Phys. Rev. Lett.\/} {\bf 100} 013001
  \urlprefix\url{https://doi.org/10.1103/PhysRevLett.100.013001}

\bibitem{Hite2012}
Hite D~A, Colombe Y, Wilson A~C, Brown K~R, Warring U, J\"ordens R, Jost J~D,
  McKay K~S, Pappas D~P, Leibfried D and Wineland D~J 2012 {\em Phys. Rev.
  Lett.\/} {\bf 109} 103001
  \urlprefix\url{https://doi.org/10.1103/PhysRevLett.109.103001}

\bibitem{Harlander2010}
Harlander M, Brownnutt M, H\"ansel W and Blatt R 2010 {\em New J. Phys.\/} {\bf
  12} 093035 \urlprefix\url{https://doi.org/10.1088/1367-2630/12/9/093035}

\bibitem{Monroe2013}
Monroe C and Kim J 2013 {\em Science\/} {\bf 339} 1164--1169
  \urlprefix\url{https://doi.org/10.1126/science.1231298}

\bibitem{Jechow2011}
Jechow A, Streed E~W, Norton B~G, Petrasiunas M~J and Kielpinski D 2011 {\em
  Opt. Lett.\/} {\bf 36} 1371--1373
  \urlprefix\url{https://doi.org/10.1364/OL.36.001371}

\bibitem{Mehta2016}
Mehta K~K, Bruzewicz C~D, McConnell R, Ram R~J, Sage J~M and Chiaverini J 2016
  {\em Nat. Nanotechnol.\/} {\bf 11} 1066--1071
  \urlprefix\url{https://doi.org/10.1038/nnano.2016.139}

\bibitem{Cetina2013}
Cetina M, Bylinskii A, Karpa L, Gangloff D, Beck K~M, Ge Y, Scholz M, Grier
  A~T, Chuang I and Vuleti\'c V 2013 {\em New J. Phys.\/} {\bf 15} 053001
  \urlprefix\url{https://doi.org/10.1088/1367-2630/15/5/053001}

\bibitem{Eltony2013}
Eltony A~M, Wang S~X, Akselrod G~M, Herskind P~F and Chuang I~L 2013 {\em Appl.
  Phys. Lett.\/} {\bf 102} 054106-- ISSN 00036951
  \urlprefix\url{https://doi.org/10.1063/1.4790843}

\bibitem{Steiner2013}
Steiner M, Meyer H~M, Deutsch C, Reichel J and K\"ohl M 2013 {\em Phys. Rev.
  Lett.\/} {\bf 110} 043003
  \urlprefix\url{https://doi.org/10.1103/PhysRevLett.110.043003}

\bibitem{Brandstaetter2013}
Brandst\"atter B, McClung A, Sch\"uppert K, Casabone B, Friebe K, Stute A,
  Schmidt P~O, Deutsch C, Reichel J, Blatt R and Northup T~E 2013 {\em Rev.
  Sci. Instrum.\/} {\bf 84} 123104
  \urlprefix\url{https://doi.org/10.1063/1.4838696}

\bibitem{Takahashi2017}
Takahashi H, Kassa E, Christoforou C and Keller M 2017 {\em Phys. Rev. A\/}
  {\bf 96} 023824 \urlprefix\url{https://doi.org/10.1103/PhysRevA.96.023824}

\bibitem{VanDevender2010}
VanDevender A~P, Colombe Y, Amini J, Leibfried D and Wineland D~J 2010 {\em
  Phys. Rev. Lett.\/} {\bf 105} 023001
  \urlprefix\url{https://doi.org/10.1103/PhysRevLett.105.023001}

\bibitem{VanRynbach2016}
Van~Rynbach A, Maunz P and Kim J 2016 {\em Appl. Phys. Lett.\/} {\bf 109}
  221108-- ISSN 00036951 \urlprefix\url{https://doi.org/10.1063/1.4970542}

\bibitem{Gross1949}
Gross B 1949 {\em J. Chem. Phys.\/} {\bf 17} 866--872 ISSN 00219606
  \urlprefix\url{https://doi.org/10.1063/1.1747079}

\bibitem{Imburgia2016}
Imburgia A, Miceli R, Sanseverino E~R, Romano P and Viola F 2016 {\em IEEE
  Trans. Dielectr. Electr. Insul.\/} {\bf 23} 3126--3142 ISSN 1070-9878
  \urlprefix\url{https://doi.org/10.1109/TDEI.2016.7736878}

\bibitem{Othman2016}
Othman N~A, Piah M~A~M and Adzis Z 2016 {\em Renew. Sustain. Energy. Rev.\/}
  {\bf 70} 413--426 ISSN 1364-0321
  \urlprefix\url{https://doi.org/10.1016/j.rser.2016.11.237}

\bibitem{Eble2010}
Eble J~F, Ulm S, Zahariev P, Schmidt-Kaler F and Singer K 2010 {\em J. Opt.
  Soc. Am. B\/} {\bf 27} A99--A104
  \urlprefix\url{https://doi.org/10.1364/JOSAB.27.000A99}

\bibitem{Xie2017}
Xie Y, Zhang X, Ou B, Chen T, Zhang J, Wu C, Wu W and Chen P 2017 {\em Phys.
  Rev. A\/} {\bf 95} 032341
  \urlprefix\url{https://doi.org/10.1103/PhysRevA.95.032341}

\bibitem{Ghosh1996}
Ghosh P~K 1996 {\em Ion traps\/} (Oxford: Clarendon Press) ISBN 9780198539957

\bibitem{Hunger2010}
Hunger D, Steinmetz T, Colombe Y, Deutsch C, H\"ansch T~W and Reichel J 2010
  {\em New J. Phys.\/} {\bf 12} 065038
  \urlprefix\url{https://doi.org/10.1088/1367-2630/12/6/065038}

\bibitem{Guggemos2015}
Guggemos M, Heinrich D, Herrera-Sancho O~A, Blatt R and Roos C~F 2015 {\em New
  J. Phys.\/} {\bf 17} 103001
  \urlprefix\url{https://doi.org/10.1088/1367-2630/17/10/103001}

\bibitem{Ott2016a}
Ott K, Garcia S, Kohlhaas R, Sch\"uppert K, Rosenbusch P, Long R and Reichel J
  2016 {\em Opt. Express\/} {\bf 24} 9839--9853
  \urlprefix\url{https://doi.org/10.1364/OE.24.009839}

\bibitem{Naegerl1998}
N\"agerl H~C, Leibfried D, Schmidt-Kaler F, Eschner J and Blatt R 1998 {\em
  Opt. Express\/} {\bf 3} 89--96
  \urlprefix\url{https://doi.org/10.1364/OE.3.000089}

\bibitem{Raizen1992}
Raizen M~G, Gilligan J~M, Bergquist J~C, Itano W~M and Wineland D~J 1992 {\em
  Phys. Rev. A\/} {\bf 45} 6493--6501
  \urlprefix\url{https://doi.org/10.1103/PhysRevA.45.6493}

\bibitem{Gulde2003}
Gulde S~T 2003 {\em Experimental Realization of Quantum Gates and the
  Deutsch-Jozsa Algorithm with Trapped $^{40}\mathrm{Ca}^{+}$ Ions\/} Ph.D.
  thesis Leopold-Franzens-Universit\"at Innsbruck
  \urlprefix\url{https://quantumoptics.at/en/publications/ph-d-theses.html}

\bibitem{Ames2019}
Ames B, Colombe Y and Blatt R 2019 {\em in Preparation\/}

\bibitem{Guerlin2010}
Guerlin C, Brion E, Esslinger T and M\o{}lmer K 2010 {\em Phys. Rev. A\/} {\bf
  82} 053832 \urlprefix\url{https://doi.org/10.1103/PhysRevA.82.053832}

\bibitem{Wade2016}
Wade A~C~J, Mattioli M and M\o{}lmer K 2016 {\em Phys. Rev. A\/} {\bf 94}
  053830 \urlprefix\url{https://doi.org/10.1103/PhysRevA.94.053830}

\bibitem{Ragg2019}
Ragg S, Decaroli C, Lutz T and Home J~P 2019 {\em Rev. Sci. Instrum.\/} {\bf
  90} 103203 ISSN 00346748 \urlprefix\url{https://doi.org/10.1063/1.5119785}

\end{thebibliography}

\end{document}